\documentclass[10pt]{article}

\usepackage{graphics}
\usepackage{graphicx}

\usepackage[a4paper, left=35mm,right=35mm,top=34mm,bottom=34mm]{geometry}
\usepackage[utf8]{inputenc}
\usepackage[T1]{fontenc}
\usepackage[english]{babel}

\usepackage{enumerate}
\usepackage{graphicx}
\usepackage{hyperref}
\hypersetup{
    colorlinks=true,
    linkcolor=blue,
    filecolor=magenta,      
    urlcolor=cyan,
}
\usepackage{listings}
\usepackage{color}

\definecolor{dkgreen}{rgb}{0,0.6,0}
\definecolor{gray}{rgb}{0.5,0.5,0.5}
\definecolor{mauve}{rgb}{0.58,0,0.82}
\lstdefinelanguage{MRGC++}{%
  language=C++,
  morekeywords={T, U, MPI_Irecv, MPI_Isend, MPI_Allreduce, MPI_Waitall, Compute, Map, abs, max, Swap, MPI_Recv_init, MPI_Send_init, MPI_Startall, Copy, Init, InitRecv, InitSend, InitAllReduce, Send, Recv, AllReduce, Finalize, InitSnapshot, Snapshot, SwitchAsync, SnapReduce, MPI_Test, MPI_Start}
}
\usepackage{mathtools,amsthm,amssymb,amsfonts}
\usepackage{algorithm}
\usepackage{algorithmic}
\makeatother
\theoremstyle{plain}

\theoremstyle{definition}

\theoremstyle{remark}

\usepackage{caption} 
\usepackage{fancyhdr}

\author{
  {\normalsize Guillaume Gbikpi-Benissan}\thanks{Ecole Centrale Paris, France
    (correspondence, frederic.magoules@hotmail.com).}
  \and
  {\normalsize Patrick Callet}\footnotemark[1]
  \and
  {\normalsize Fr\'ed\'eric Magoul\`es}\footnotemark[1]
}
\title{Spectral domain decomposition method for physically-based rendering of photochromic/electrochromic glass windows}
\date{}

\begin{document}
\maketitle
\thispagestyle{fancy}

\begin{abstract}
\noindent This paper covers the time consuming issues intrinsic to physically-based image rendering algorithms. First, glass materials optical properties were measured on samples of real glasses and other objects materials inside an hotel room were characterized by deducing spectral data from multiple trichromatic images. We then present the rendering model and ray-tracing algorithm implemented in Virtuelium, an open source software. In order to accelerate the computation of the interactions between light rays and objects, the ray-tracing algorithm is parallelized by means of domain decomposition method techniques. Numerical experiments show that the speedups obtained with classical parallelization techniques are significantly less significant than those achieved with parallel domain decomposition methods.
\end{abstract}


\section{Introduction}

Nowadays, virtual reality is a powerful tool to design and preview real world manufacturing. This often implies to make up 3D computer-aided models displaying objects inside an environment, and to evaluate some particular behaviors according to some predefined properties. The scope of this study is to simulate the visual aspect of a room, depending on natural lighting and materials optical properties, and in particular glasses. The complexity of the interaction between light and materials is quite hard to entirely reproduce because of the heterogeneity of their optical behavior. Therefore, physically-based rendering of the visual aspect of a lighted scene can not be done by means of a simple trichromatic description of objects color. Instead, a model including spectral properties allows to simulate, for instance, the nuances induced by the photochromic adaptation of glasses. Commonly used physically-based rendering engines~\cite{Pharr:2010:PBR:1854996}, \cite{LuxRender}, \cite{Shirley:2012:BPR:2407783.2407785}, are designed about the only extrinsic properties of materials, such as spectral reflectance and transmittance. A more complete approach additionally takes into account optical constants which are intrinsic properties. That is the case inside Virtuelium, the open source rendering software on which we experienced our study.

Beside the accuracy of a physical model of interactions between light and objects, a higher degree of computational complexity is introduced. As a consequence, ensuring fast simulations requires to compromise on the fineness of the rendering processes, hence on the quality of the resulting images. That is a particular concern in the context of real-time application where high level rendering is difficult to achieve. Parallel computing could be a privileged way to overcome the inherent time consuming issue of the physically-based rendering. Basic parallelization techniques, like splitting the set of pixels to be processed, raise a certain amount of efficiency problems. While dynamic load-balancing could mitigate the drawback related to the nonuniform distribution of objects and light sources in the whole scene, data size remains a limitation, principally due to data replication. With Domain Decomposition Methods (DDM)~\cite{SBG1996}, \cite{QV1999}, \cite{TW2005}, \cite{Jar2007}, we benefit by data parallelism principles and techniques of information sharing based on interface conditions~\cite{magoules:journal-auth:16}. In~\cite{magoules:patent:2011}, a ray-tracing domain decomposition method was proposed for accelerating the simulation of the propagation of acoustic waves. Given the good speedup results presented in~\cite{6636420}, our proposal is to study the impact of the same approach in the context of physically-based image rendering.

In the following we first describe the optical properties considered for our simulation, and the specific methodology of acquisition on glass samples and trichromatic images depicting the 3D scene. Then, after the global rendering equation, we present rendering algorithms including those used in Virtuelium, mainly based on ray-tracing. Third, our parallel domain decomposition method for accelerating these algorithms is given. At last, we close our demonstration with a discussion about the speedups we obtained, and an example of rendered image.

\section{Measurement of optical properties}

Our study mainly focused on the effects of glass materials. With samples of glass, we can acquire reflection and transmission properties by means of spectrophotometry~\cite{bass1995handbook}. More precisely, for glass materials, this technique allows to obtain a Bidirectional Transmittance Distribution Function (BTDF). The process consists in determining the spectral response of a lighted material by repeating, for a known emission spectrum and for several points on the surface of the object, measurements of energetic quantities along a wavelength range, while varying both incident and view angles. Intrinsic properties called optical constants can be determined by a technique based on spectroscopic ellipsometry~\cite{palik1985handbook, callet1998couleur}. It offers a way to measure the complex index of refraction defined by
\begin{equation}
\label{eq:optical}
\tilde{n}(\lambda) = n(\lambda)+ i k(\lambda) = n(\lambda) (1 + i \kappa(\lambda))
\end{equation}
The index of refraction, according to a wavelength $\lambda$, depends on the optical index $n(\lambda)$ and on the index of absorption $\kappa(\lambda)$. Unlike extrinsic properties which only define a spectral response, optical constants represent the electronic behavior of dielectric materials. Yet, the characterization by optical constants is more adapted to materials satisfying Fresnel conditions of non-scattering and homogeneity. Metallic surfaces, for instance, are often described by means of optical constants for the simulation of their visual appearance~\cite{cgaBerger, Woollam199444, 5111815}.

The other kinds of materials inside the room scene were described by a completely different method, as there were no real samples of the objects to be represented. Indications about these materials were given by a set of images depicting their trichromatic appearance under various lighting conditions. There is no exact method to deduce spectral information from given trichromatic images, due to the fact that a given RGB value can be produced by infinity of different spectra. However, the Matrice-R theory, defined by Cohen and Kappauf in 1982~\cite{cohen1982metameric}, is a possible solution to achieve accurate RGB-to-spectrum conversion. According to this theory, every spectrum can be decomposed into two components:
\begin{itemize}
\item the fundamental function that is unique and contains the color stimulus of the spectrum,
\item a metameric black function that gives X=Y=Z=0 when converting to CIE XYZ.
\end{itemize}
An infinite number of metameric black functions exists. In term of calculation, the metameric black space is orthogonal to the color stimulus space. For this reason, Cohen defined a matrix equation we can use to compute the fundamental function from every RGB values. The practical accuracy of this method could be evaluated by computing spectra from a set of representative RGB values, then reconverting these spectra into RGB space, for qualitative comparison.

\section{Rendering model}

Our rendering equation is deduced from the Radiative Transfer Equation (RTE) and described by Kajiya~\cite{Kajiya:1986:RE:15886.15902} as follows:
\begin{equation}
\label{eq:radiative_theo}
L_{r}(\vec{\omega_{o}}) = \int_{\Omega} F_{r}(\vec{\omega_{i}}, \vec{\omega_{o}}) L_{i}(\vec{\omega_{i}}) \vec{n}.\vec{w_{i}} d \omega_{i}
\end{equation}
where $\Omega$ is a dome of incident lights. Knowing the Bidirectional Reflectance Distribution Function (BRDF), $F_{r}$, a remitted light $L_{r}$, in a direction $\vec{\omega_{o}}$, is computed from every incident light $L_{i}$ emitted by the dome. The reasoning for the BTDF is the same except that we do not only consider a dome for incident lights but an entire sphere since the studied surface is non-opaque. On an applicative level, we only consider point or directional light sources, hence we simplify the equation (\ref{eq:radiative_theo}) as follows:
\begin{equation}
\label{eq:radiative_appli}
L_{r}(\vec{\omega_{o}}) = \sum_{s 1}^N F_{r}(\vec{\omega_{s}}, \vec{\omega_{o}}) L_{s}(\vec{\omega_{s}}) \vec{n}.\vec{w_{s}}
\end{equation}
where $N$ is the number of light sources, $L_{s}$, the emission spectrum of the current light source $s$ and $\vec{\omega_{s}}$, the incident direction.

Naturally, given that objects inside a 3D scene reflect a part of the light emitted from light sources, these objects should all be responsible for the illumination of the whole scene. That is what we call Global Illumination (GI). However, such a precise computation is sometimes avoided, as it is much simpler and faster to estimate only interaction between objects and light sources. That is local illumination. Our rendering software, Virtuelium, observes both local and global illuminations through two algorithms of each kind, respectively the ``Scanline rendering" and the ``Photon Mapping".

\section{Rendering algorithms}

\paragraph{Local illumination}
The ``Scanline rendering"~\cite{Wylie:1967:HPD:1465611.1465619} is based on the inverse ray tracing algorithm~\cite{Arvo86backwardray}. Given that the image to be computed can be viewed as a matrix of pixels, a light ray is emitted from each pixel, orthogonally to the image plane. When a ray intersects the closest object on its path, we have to evaluate the received luminance at the given viewed direction. For that, new rays are shot from the hit point toward each light source, thus determining all the needed incident directions. Then, secondary rays are thrown regarding to reflection and/or refraction laws and the process is repeated. The algorithm stops when the energetic value of the ray goes bellow a threshold, or after the ray has bounced a predetermined number of times. The main difference between ray tracing algorithms lies in the way polygons of objects are sorted. In ``Scanline rendering", every polygons are projected onto the image plane. Then, the image is computed line by line, from top to bottom, determining the color of each pixel by considering the closest polygons around. Another very common algorithm is the Z-buffer technique~\cite{Catmull:1974:SAC:907242} which is nowadays implemented by default on graphic cards. The main advantage of the ``Scanline rendering" is that each pixel is evaluated only once. In return, the memory cost is high because all the polygons of the scene must be loaded at the same time, leading to bad performances for scenes with complex geometries.

\paragraph{Global illumination}
GI is a major progress in the quest of photo-realism, and a lot of very different techniques have been developed. ``Radiosity" methods~\cite{Wallace:1987:TSR:37402.37438, Sillion:1989:GTM:74333.74368} transform the phenomenon into a system of linear equations, solved either by direct method~\cite{journals:vc:BuD89} (very effective but with a high complexity), or by iterative algorithms~\cite{Cohen:1988:PRA:54852.378487}. In another direction, the stochastic algorithm of ``Monte-Carlo"~\cite{134595} is sometimes used despite its slower convergence. ``Path Tracing" methods~\cite{CGF:CGF1863} launch random rays from pixels of the image plane until one hits an object. It can be bi-directional (rays are shot from camera and sources simultaneously). ``Metropolis Light Transport" algorithm (MTL)~\cite{Hachisuka:2008:PPM:1409060.1409083} optimizes ``Path Tracing" by replacing the random shooting by heuristics. The ``Photon Mapping" algorithm implemented in Virtuelium was first defined by Jensen in 1996, and is improved since this date~\cite{Jensen:1996:GIU:275458.275461, Jensen:2004:PGG:1103900.1103920}. It decomposes the rendering process into two steps which are executed sequentially. In the pre-rendering step, the position of photons (light rays launched from a light source) hitting objects are stored in several appropriate structures (photon maps). At least, two photon maps are needed, one for the global illumination itself and one for caustics. Then, a next step consists in evaluating four different contributions based on the fact that $L_{r}(\vec{\omega_{o}})$ can be decomposed into a sum of different integrals. First, the direct and specular contributions are computed the same way than in ``Scanline Rendering". Then, the caustic and indirect diffuse contributions are deduced from the two photon maps. Unlike this version of ``Photon Maping", most recent versions are progressive~\cite{Hachisuka:2008:PPM:1409060.1409083, Hachisuka:2011:RAP:2019627.2019633}.

\section{Parallel computing}

Commonly, a parallel image rendering algorithm decomposes the image grid, such that each pixel can be treated separately without any interaction. However,  because of the heterogeneous spacing of objects, materials and lights-sources in the scene, it could be longer to compute some area of the image. Thus, a dynamic job-balancing mechanism is required to ensure that faster threads work more and that there is no inactivity period for any of them. The same idea can be applied to the set of light sources, to the light rays, or even to the spectral data when dealing with a full spectral rendering. But such a distribution requires to replicate the whole scene geometry onto each computational node. Indeed, on one hand, predetermining the whole light path of a ray is nearly impossible, and on the other hand, each polygons in the scene can be hit several times by different rays. Shared memory can be used to assure that only a single copy exists on a computational node but the problem remains on multiple-node architectures. Thus, we propose to apply the ray-tracing domain decomposition method introduced in~\cite{magoules:patent:2011}.

By splitting a global domain into several small sub-domains, domain decomposition methods~\cite{SBG1996}, \cite{QV1999}, \cite{TW2005}, \cite{Jar2007} allow to load input data and to gather results in a parallel way, as each sub-domain can be associated to a unique processor. The method described in~\cite{magoules:patent:2011} takes advantage of some efficient domain decomposition techniques~\cite{magoules:journal-auth:4}, \cite{magoules:journal-auth:16}, \cite{magoules:journal-auth:21}. Besides the splitting of the global geometry itself, information along interfaces is shared between computational units which are processing neighboring sub-domains~\cite{magoules:journal-auth:3}, \cite{magoules:journal-auth:2}. A continuous approach~\cite{Des1993}, \cite{Gha1997}, \cite{CN1998}, \cite{magoules:journal-auth:28}, \cite{magoules:journal-auth:23}, \cite{magoules:journal-auth:18}, \cite{magoules:journal-auth:14}
can be used to design efficient interface conditions.
Similarly a discrete approach~\cite{magoules:journal-auth:8}, \cite{magoules:proceedings-auth:6}, \cite{magoules:journal-auth:29}, \cite{magoules:journal-auth:12}, \cite{magoules:journal-auth:20} 
can be used, which may increase significantly the performance of the algorithm.
The link between the continuous and discrete interface condition can be established like in~\cite{magoules:journal-auth:17}.

The method used in this work is based on the domain decomposition methods \cite{magoules:journal-auth:24}, \cite{magoules:journal-auth:9}, \cite{magoules:journal-auth:10}, \cite{magoules:journal-auth:13}, where here the interface conditions assure the continuity of the light ray properties (such as direction, amplitude, angle, etc.) from one sub-domain to another one. Yet, unlike classical domain decomposition methods, a computational unit does not process only one sub-domain. In order to cover load balancing issues, each processor starts by loading a certain number of sub-domains, according to memory limitation. When it remains few light rays to be handled in a sub-domain, this sub-domain is unloaded if there is still other currently not handled sub-domains with a lot of rays not processed. Then the processor starts loading one or more of these sub-domains while handling another sub-domain already available in the memory. Unloading sub-domains allows doing most of the results gathering during the processing of other rays. This overlapping gathering and processing is efficient since gathering mainly uses the communication system. A more complete description of an efficient implementation can be found in~\cite{magoules:patent:2011}.

\section{Results and discussions}

\begin{figure}
\centering
\scalebox{1.0}{\includegraphics[width=0.45\textwidth]{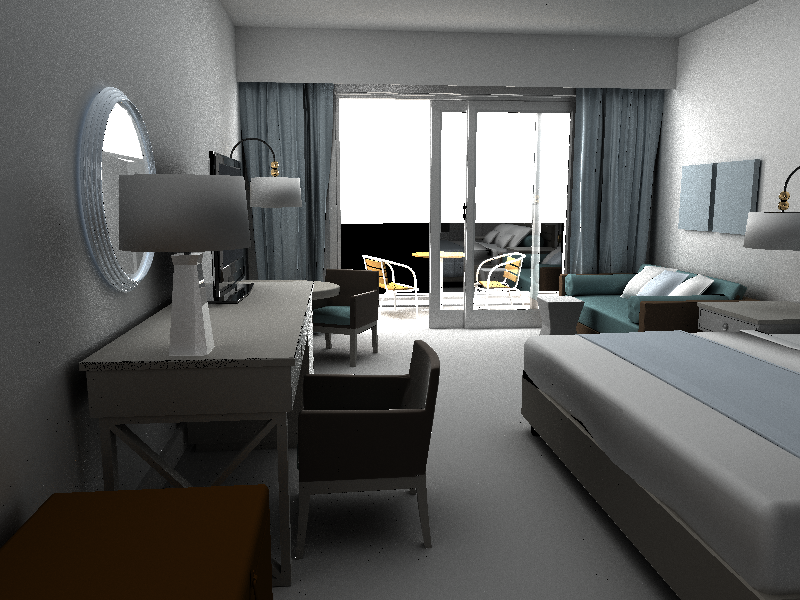}}
\caption{Illustration of the image rendering of the interior of a hotel room.}
\label{fig:virtuelium01}
\end{figure}

An image rendered with Virtuelium is presented in Figure~\ref{fig:virtuelium01}. The presented scene of a hotel room is visually a simplification but the glasses behavior, and the role of filter the windows are playing for the sunlight, are very accurate because of the direct use of Fresnel definition. Glasses are defined by using Fresnel indices of borosilicate glass Schott (BK7). Replacing these transparent glasses by colored glasses in Virtuelium simply means adapting the Fresnel definition of the material.

\begin{table}
\caption{Speedup of the Virtuelium DDM program (Ethernet) with respect to the number of threads and sub-domains.}
\label{tab:ddm_virtuelium}
\centering
{\small
\begin{tabular}{|l|c|c|c|c|}
\hline 
 & 16 & 32 & 64 & 128 \\
 & threads & threads & threads & threads \\
\hline %
{1 sub-domain}  & 10.6 & 16.7 & 25.4 & 20.2 \\
{2 sub-domains} & 11.9 & 22.1 & 34.3 & 45.4 \\
{4 sub-domains} & 10.4 & 22.3 & 35.1 & 50.7 \\
{8 sub-domains} & 11.2 & 24.2 & 39.8 & 66.9 \\
\hline 
\end{tabular}
}
\end{table}

Speedups of Virtuelium execution are shown in Table~\ref{tab:ddm_virtuelium}. They are very closed to results
presented in~\cite{6636420} for the beam-tracing acoustic simulation software. Simulations were run on a hybrid, both distributed and shared memory, computational platform consisting of 4 nodes containing a quad core processor (a total of 16 cores). Each node were provided with 8 Gigabytes RAM (Random Access Memory). As we were expecting, DDM techniques significantly improved the performance of the parallelization. Although the speedups from acoustic simulation were quite better~\cite{6636420}, we can notice that in both cases, from 16 to 128 threads, 8 sub-domains decomposition allowed to multiply the speedup by nearly 6, while classical parallelization only reach a factor less than 2. On another hand, for a fixed number of threads, the speedup keeps increasing as the number of sub-domains do.

\section{Conclusion}

In this paper, we proposed an original ray-tracing domain decomposition method for image rendering with natural lighting. According to domain decomposition methods principle, light rays characteristics have been matched as interface constraints between neighboring sub-domains. We presented a test case on a model of a hotel room where we particularly deal with glass material properties. It outlined the performance and efficiency of our method, relatively to multi-core architectures.

\section*{Aknowledgements}

The authors acknowledge partial financial support from the Callisto-Sari project of the P\^ole de Comp\'etitivit\'e Advancity, Ville et Mobilit\'e Durables, Cap Digital Paris Region, France.

\bibliography{ref}
\bibliographystyle{abbrv}

\end{document}